\documentclass[aps,amsmath,amssymb,prl,twocolumn,showpacs,floatfix,tightenlines]{revtex4}
\usepackage{bm,natbib}
\usepackage{graphicx}

\begin{document}

\title{The specific heat jump at the superconducting transition and the quantum critical nature of the normal state of Pnictide superconductors.}
\author{J. Zaanen}

\affiliation{Instituut-Lorentz for Theoretical Physics, Universiteit Leiden, P. O. Box 9506, 2300 RA Leiden, The Netherlands}

\begin{abstract}
Recently it was discovered that the jump in the specific heat at the superconducting transition in pnictide superconductors is
proportional to the superconducting transition temperature to the third power, with the superconducting transition temperature
varying from 2 to 25 Kelvin including underdoped and overdoped cases. Relying on standard scaling notions for the thermodynamics of
 strongly interacting quantum critical states, it is pointed out that this behavior is consistent with a normal state that is a quantum critical metal 
 undergoing a pairing instability.    
\end{abstract}

\date{\today}

 \maketitle

At present it is widely believed that the 'high' Tc superconductivity observed in pnictide superconductors\cite{hosona}  is explained by 
the classic Bardeen-Cooper-Schrieffer (BCS) mechanism with the caviat that the pairing
glue is likely not phonons but instead related to magnetic fluctuations. There is abundant evidence for the opening of a gap
in the spectrum of electronic excitations at the superconducting transition temperature ($T_c$) as
related to the binding of electrons in Cooper pairs. However, for the BCS formalism  to be valid (including the
non-phonon glue possibility) one also has to assume that the normal state is a Fermi-liquid.
$T_c$ is after all determined by the balance of the free energy of the superconducting- and normal states and one has
to understand the nature of both states in order to find out why the transition happens at a particular temperature. More
specifically, the pairing instability  in classic BCS  is governed by the electronic pair susceptibility 
which is a four point linear response function. It is a specialty of the Fermi-liquid that the information on
the pair susceptibility is entirely contained in the single fermion response, but this will a-priori not be the case in any form of 
non-Fermi liquid matter\cite{she}. This issue is recognized both in the context of  the 'quantum critical' heavy fermion superconductors\cite{heavyf}
and the
optimally doped cuprate superconductors\cite{anderson}  where it is well established that the normal states are poorly understood non-Fermi liquids.  
The situation in the pnictides is less clear. Although their normal states are bad metals showing quite large, strongly
temperature dependent resistivities and other anomalous transport properties,  it can be argued that this is just caused by the 
low carrier density\cite{basov}.  However, the normal state is still rather poorly characterized empirically and it has been hypothesized that
it might be quantum critical, perhaps driven by the vanishing of the antiferromagnetism and/or structural phase transition of the parent compounds  
under influence of doping\cite{Si}. 

The thermodynamics of strongly interacting quantum critical states is governed by simple scaling behaviors,
that are applicable also when a microscopic understanding of the critical state
is completely absent\cite{roschsi,babak}. Here I want to draw attention to recent  measurements, revealing a surprising scaling of 
the jump of the specific heat at the superconducting transition versus $T_c$  in the 122 pnictide  family, involving the full doping range where 
superconductivity occurs. In Fig. 1 I reproduce the results by  Bud'ko, Ni and Canfield ('BNC')\cite{budko}, and include newer data by Mu {\em et al.}\cite{wen},
revealing that the specific heat jump shows a scaling behavior  $\Delta C_p = A T_c^3$, where $A$ is a constant, over a 
dynamical range of more than a decade with $T_c$ varying between 3 and 35 $K$. To explain such a scaling behavior within the realms
of conventional BCS theory one needs extreme fine tuning. Here I want to point out that this scaling finds a natural explanation in terms of 
the normal state being in some fermionic quantum critical phase that undergoes a pairing instability. 

\begin{figure} 
\begin {center}
\includegraphics[width=0.48\textwidth]{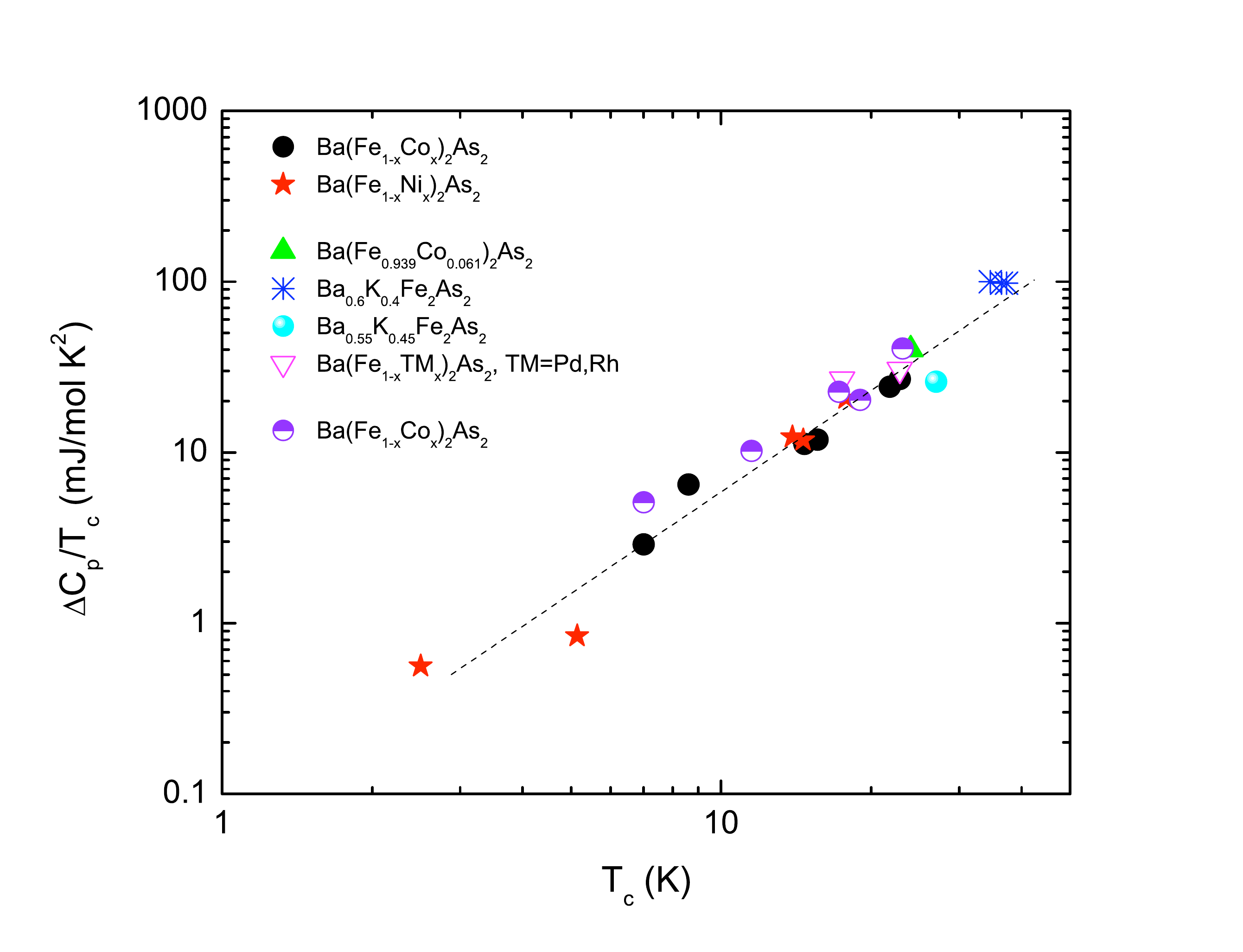}
\end {center}
\caption{The scaling behavior of the ratio specific jump at the superconducting transition  and the superconducting transition temperature $T_c$
 $\Delta C_p /T_c$ 
versus $T_c$ in the 122 pnictides, as reproduced from  Bud'ko {\em at al.}\cite{budko}, with the independent
results by Mu {\em et al.}\cite{wen} (purple half-filled circles) analyzed in a similar fashion and added. }
 \label {Figbudko}
\end{figure}

Let us first consider the problems of principle one encounters rationalizing Fig. 1 in terms of conventional Fermi-liquid based pairing theory.
The jump in the specific heat at the transition  finds its origin in the fact that the superconducting gap opens up exponentially fast. The specific 
heat just above the transition reflects the number of degrees of freedom that contributes to the entropy at the temperature $\simeq T_c$ and since 
these are determined by the renormalized Fermi-energy of the metal the Sommerfeld expression  $C_p = \gamma T$ determines 
the specific heat,  where 
$\gamma = N_0 \sim 1/E_F$.  When the SC gap opens, these degrees of freedom suddenly disappear from the energy window $\simeq T_c$ 
and therefore the specific heat jumps by an amount  $\Delta C_p = B C_p^m (T_c) = B \gamma T_c$ where $B$ is a constant of order unity
depending on the details of the thermal evolution of the gap (in weak coupling s-wave BCS,  $B=1.14$) . Therefore, the ratio $\Delta C_p/(k_B T_c) 
\simeq N_0$, the density of states in the metal at the temperature $T_c$.  Although numerical factors do depend on complicating factors like
multi-gap superconductivity, strong coupling effects and so forth, the scaling of the jump with temperature will not change since it is governed by the
Fermi-energy, the largest scale of the Fermi-liquid.   Within this conventional interpretational framework, the 'BNC scaling'
revealed by Fig. 1 would imply that the density of states in the metal would actually vary like $T_c^2$. As Bud'ko {\em et al.} argue, this is quite hard
to understand because a Fermi-liquid normal state would imply that LDA band structure should at least yield a qualitative impression 
of the density of states of the metal in the doping range of the superconducting dome. These calculations however indicate that the density 
of states should evolve quite smoothly\cite{ldacal}. In fact, one has to deal with a severe 'naturalness' problem relying on the conventional BCS
interpretation. Internal consistency requires that the coupling constant that determines $T_c$ is itself determined by the density of states of the metal: 
$\lambda = N_0 V$ where $V$ is the strength of the glue mediated attractive interaction. 
The problem can be directly inferred from the weak coupling BCS
expression for the transition temperature $k_BT_c \simeq \hbar \omega_0 exp (-1/\lambda)$. Writing $N_0 = C_0 T^2_c$, and inserting 
$\lambda = V C_0 T^2_c$ one finds the condition,
\begin{equation}
T^2_c \ln (\frac {\hbar \omega_0} {k_B T_c}) \simeq \frac{1}{C_0 V}
\label{BCSfail}
\end{equation}
Assuming  that the glue retardation scale $\hbar \omega_B$ is doping independent $1/V$ should vary
precisely like $T^2_c \ln (\hbar \omega_B / k_B T_c)$ over a range where $T_c$ varies by more than an order of magnitude.
Alternatively, assuming it is due to the retardation scale one has to require that $\hbar \omega_0 \simeq k_B T_c \exp (1/ VC_0 T^2_c)$:
these are most unnatural fine tuning conditions indeed! 
This fine tuning  problem becomes only worse using more fanciful expressions like the McMillan formula. As a limiting
case, consider the ultra strong coupling case of Dynes and Allen\cite{Dynes} where $T_c = 0.183 \sqrt{ \lambda \langle \omega_0^2 \rangle}$;
this would turn into the extreme fine tuning  condition $1 = 0.183 \sqrt{ C_0 V   \langle \omega_0^2 \rangle}$.

This paradox finds its origin in the assumption that the normal state is a Fermi-liquid.  The Fermi-liquid is exceptional in the regard 
that everything is eventually governed by the scale of the Fermi-energy. This is obvious for the specific heat, but it is also underlying the standard BCS
theory. Staying within the realms of a pairing instability, the superconducting transition is governed by the  criterium 
$1- V \chi'_{pp} (\omega =0, \vec{q} =0) =0$, where $V$ is the interaction strength and $\chi'_{pp}$ is  the zero frequency, zero momentum
real part of the electronic pair susceptibility. For the special case of non-interaction fermions this susceptibility becomes marginal in a scaling sense\cite{she}.
The imaginary part is independent of frequency and its magnitude is therefore set by $1/ E_F$. By accounting for retardation via the Kramers-Kronig
transform\cite{she}  $\chi' (\omega=0) = \int _0^{2\omega_B} \chi'' ( \omega) /\omega d\omega= N_0 \int_0^{2\omega_B} d \omega / \omega$ one recovers the 
'BCS logarithm' that is responsible for the exponential dependence of the gap and $T_c$ on the coupling constant. The relation between the 
density of states measured by the specific heat jump and the coupling constant as of relevance to $T_c$ is therefore unique for the Fermi-liquid:
for {\em any} other fluid fermion state  there will not be a direct relation between these two quantities. The apparent paradox of the previous
paragraph can therefore be seen as strong evidence that the normal state of the pnictides is not a Fermi-liquid.  

Let me now discuss why the scaling of Fig. 1 is suggestive of a quantum critical state. In fact, besides standard scaling arguments one just needs
that the system behaves BCS like in the sense that a pairing gap rapidly opens in the spectrum of, now quantum critical, electronic 
excitations at $T_c$. This is phenomenologically implied  by the very fact that the specific heat jumps. Therefore, the jump measures the normal 
state specific heat  at $T_c$, associated with the electrons that pair up in the superconductor: $C_p (T_c) \simeq A_C T_c^3$.  It appears 
that the only way one can explain this scaling behavior without running into other fine-tuning issues is by asserting that the specific heat in the
metallic state over the  whole superconducting range has a 'universal' form $C_p = A'_C T^3$, being just probed at different temperatures (the
$T_c$'s).  As I will discuss in more detail, this has far reaching and unexpected consequences, and a direct experimental check of this assumption
would be desirable. However, it might well appear to be experimentally impossible to disentangle an electronic specific heat from a phonon 
background with the same $T^3$ temperature dependence.     

This $T^3$ specific heat is in turn is a rather famous property of the thermodynamics of a strongly interacting quantum critical system\cite{blackh}. Dealing with a scale invariant ('conformal') quantum system  one learns from thermal field theory that at finite temperature the scale invariance is broken by the finite radius of the imaginary time circle 
$R_{\tau} = \hbar / (k_B T)$. When the fixed point is strongly interacting (obeying hyperscaling)  the singular part of the free energy acquires the scaling form\cite{roschsi,babak},
\begin{equation}
F_s = - \rho_0 \left( \frac{T}{T_0}  \right)^{(d+z)/z} f \left( \frac{r}{(T/T_0)^{y_r/z)}} \right)
\label{freeen}
\end{equation}
where $d$ and $z$ are the number of space dimensions and the dynamical critical exponent, respectively, while $T_0$ is the high energy cut-off.
The cross-over function $f$ is governed by the zero temperature coupling constant $r$ with scaling dimension $y_r$ and since there is no
singularity at $r=0, T >0$ it expands as $f(x \rightarrow 0) = f(0) + x f'(0) + \cdots$. Since the specific heat $C_p = -T (\partial^2 F/\partial T^2)$ it
follows\cite{babak},
\begin{equation}
C_p = A_{cr}  \left( \frac{T}{T_0} \right)^{d/z}
\label{specheat}
\end{equation}
where $A_{cr} = \rho_0 f(0)(d+z)d/z^2$. 
The specialty of the specific heat of a strongly interacting quantum critical system is the fact that its temperature is governed by the engineering
dimensions $d$ and $z$, as rooted in the finite size scaling . In pnictides it is reasonable to take $d=3$ and consistency with  the BNC scaling 
suggests that $z=1$ reflecting an 'emergent Lorentz invariance'.  Notice also that it requires that non-singular 
contributions to the electronic free energy are absent. This is not unreasonable given that we are dealing with fermionic quantum critical matter:
it is hard to reconcile fermion statistic with the notion that some electrons stay in a Fermi-liquid and others go critical -- electrons are after all
indistinguishable.     

If the above makes sense,  we are likely dealing with some unknown form of fermionic quantum criticality and it is a-priori impossible to make 
definitive statements regarding the constant $A'_C$.  In 1+1 D it is set by the central charge of the 2D conformal field theory, but this is much less
understood in higher dimensions.   The only example where its magnitude for a strongly
interacting quantum critical state in higher dimensions is known is  the maximally supersymmetric Yang-Mills theory in the large
 $N$ limit at zero chemical potential. Its thermodynamics
is according to the string theoretical AdS/CFT correspondence governed by Black hole thermodynamics in an Anti-de-Sitter space time with one
extra dimension\cite{blackh,polchin}. As in the previous paragraph, the $T^d$ temperature dependence is fixed by scaling but it turns out that the specific heat in the large $N$ limit is 
just $3/4$ of the 'Debye' specific heat associated with $N$ free fields,   
\begin{equation}
C_p \simeq \frac{4 \pi^4} {15} R  N^2 \left( \frac{T}{T_0} \right)^3
\label{specheatqu}
\end{equation}
To give an impression of the numbers in the game, I estimate the prefactor $A'_{C}$ from the data in ref.\cite{budko} to be $\simeq 26 mJ/mol K^4$. Assuming
$N=3$, this just becomes the Debye specific heat for phonons in $d=3$ and the UV cut-off ("Debye")  temperature becomes $T_0 =  900 K$.
Alternatively, taking $N = 1$ it follows that $T_0 = 432 K$.  It is unlikely that the pnictide  critical metal has any direct dealings with this zero
density large $N$ gauge theory, but this example illustrates that the gross magnitude of  $A'_C$ is in first instance determined by the UV
cut-off scale which appears to fall in a reasonable regime for the electron system under consideration.

What is the relationship between the specific heat and the superconducting transition temperature when the normal state is a quantum  
critical metal? On general grounds one expects that the fine tuning problems encountered in the Fermi-liquid case disappear, since 
the 'Fermi energy as common denominator' for specific heat and $T_c$ is no longer a factor. This can be illustrated using the simple
scaling theory for 'BCS' pairing in a quantum critical normal state as  recently discussed by She and myself\cite{she}. This departs from 
the assumption that a truly conformal fermionic state is perturbed by an external retarded bosonic mode causing attractive interactions, 
with the consequence that the BCS gap equation is still valid. The information on the fermion system enters through the fermionic 
pair susceptibility, with a form that  is fixed by the conformal invariance and parametrized by an anomalous dimension $\eta_p$ and dynamical
critical exponent $z$. The 
superconducting transition temperature is now determined by\cite{she},
\begin{equation}
\label{qcgap}
k_B T_c  \simeq  \hbar \omega_0  \left({1+\frac{1}{\tilde{\lambda}}\left(\frac{2\omega_0}{T_0} \right)^{\frac{2-\eta_p}{z}} } \right) ^{-\frac{z}{2-\eta_p}},
\end{equation}  
where $\omega_0$ and $\tilde{\lambda}$ represent the glue frequency and pairing strength, respectively, The UV cut-off scale $T_0$ also enters through the normalization of the dimensionless coupling $\tilde{\lambda} \sim V/T_0$ where $V$ is the dimensionful coupling.  By varying $V$ 
and/or $\omega_0$ one can vary $T_c$ at will, while the specific heat jump automatically tracks the BNC scaling. The fine tuning problem of
conventional BCS has completely disappeared.    
   
Without claiming it to be an unique explanation, the scenario in the above is at least consistent with the strong constraints posed by the BNC scaling.  
It does have however a quite surprising and far reaching consequence for the physics of the pnictides.  It suggests that the normal state is a 
quantum critical {\em phase} extending over the whole superconducting doping range: this scenario revolves around the notion that there is a
metal phase with a specfic heat that is doping independent.  The prevailing view is that when quantum criticality
is relevant for pnictides, it should be tied to the isolated quantum critical point (QCP) associated with the disappearance of the magnetism and/or
lattice distortion. After all, this QCP seems at least in the 112 system coincident with the doping level where $T_c$ is maximal\cite{ni,chu,canfield,ni1}.
As well documented in the heavy fermion systems, a quantum critical metal 'fan' as function of increasing temperature or 
energy is centered at such a zero temperature QCP. In the cuprates the situation is less clear\cite{hussey}, but a similar analysis as presented 
here indicates that there is certainly not a 'universal'  quantum critical thermodynamics\cite{babak}.  However, the BNC scaling appears to be
inconsistent with a metallic state that is controlled by an isolated QCP on the doping axis. The expectation would be that in a doping regime 
close to optimal doping the transition would go directly from the quantum critical metal to the superconductor, but farther out in the 'wings' of
the superconducting dome the metal would first crossover to a stable, scale-full state with the transition to the superconductor happening
at  lower temperature. The emergence of such a scale (Fermi energy, pseudogap, whatever) should show up as a failure of the BNC scaling 
when $T_c$'s become low. A loophole is that the crossover temperatures might increase very slowly in moving away from the QCP. However, this 
argument excludes the magnetic quantum critical point as the cause of the quantum criticality. The thermal transition to the finite temperature 
magnetic order can only happen at a temperature below the quantum critical crossover temperature, and Fig. 1 contains a number of points
at doping levels where the superconducting transition is (much) lower than the antiferromagnetic transition.  This does not imply
that the magnetic/structural  QCP is irrelevant for the superconductivity. It might well be that, as in the Fermi-liquid, the critical fluctuations
of the bosonic order parameter are a  source of strong retarded attractive interactions also in the quantum critical metal\cite{chubukov}.

In conclusion, thermodynamics is a powerful source of information dealing with quantum critical states of matter since it is subjected 
to strong scaling principles that makes it possible to arrive at  phenomenological insights even when a more microscopic understanding
is completely absent. If the present claim based on thermodynamics is correct that pnictide metals are quantum critical, this should have 
far reaching ramifications for other experiments on the normal state. It is hoped that this work will form a source of inspiration for a 
concerted effort to study this normal state in much further detail.   

\acknowledgements  
{\em Acknowledgements.}
I would like to thank P.C. Canfield and S.L. Bud'ko for providing figure 1. This work was 
carried out during the programs on higher Tc superconductivity and AdS/CMT at the Kavli Institute for Theoretical Physics
and it has much profited from interactions with participants in these programs, with a special mention of P.C. Canfield, S. Sachdev, E. Fradkin
and T.M. Rice. This work was concieved during the pnictide program at the Kavli Institute for Theoretical Physics China, and
I acknowledge in particular discussions with Z.Y. Weng and H.H. Wen. This work is supported by the 
Nederlandse Organisatie voor Wetenschappelijk Onderzoek (NWO) via a Spinoza grant, and by the National Science
foundation under Grant No. PHY05-51164.

\end{document}